\documentclass[epjCONF,columns,a4paper]{svjour}
\usepackage{graphics}
\usepackage[varg]{txfonts} 
\usepackage[latin1]{inputenc}
\session-title{Hadron Collider Physics symposium (HCP-2011)}
\newcommand{\pt}{\ensuremath{p_{\rm{T}}}}

\begin{document}
\title{Heavy flavour results from ATLAS}
\author{P. J. Bell\thanks{\email{paul.bell@cern.ch}}}
\institute{Universit\'e de Gen\`eve, 1211 Gen\`eve 4, Switzerland}
\abstract{
A selection of heavy-flavour physics results from the ATLAS experiment is presented, based on data collected 
in proton-proton collisions at the LHC during 2010.
Differential cross-sections for the production of heavy flavours, charmonium and bottomonium states 
and $D$-mesons are presented and compared to various theoretical models. Results of $B$-hadron lifetime
 measurements are also reported.
} 

\maketitle
%


\section{Introduction}
\label{intro}

The goals of the heavy-flavour physics programme at ATLAS are to test theoretical models for 
heavy-flavour production within the Standard Model (SM) and to search for new physics through rare decays or 
new sources of CP violation. These proceedings present a non-exhaustive selection of analyses completed 
during 2011 based on 2010 data, divided into cross-section measurements (Section 2) and lifetime measurements
(Section 3).

Details of the ATLAS detector may be found in~\cite{bib:atlas}. The sub-detectors of greatest
importance to the analyses presented here are in the Inner Detector (ID) tracker and Muon Spectrometer (MS)
systems. In addition, in many cases the data collection has relied on specific $B$-physics trigger selections 
implemented in the Higher Level Trigger (HLT).

\section{Cross-section measurements}

\subsection{Heavy-flavour cross-section measurements}
\label{sec:1}

At low transverse momentum ($\pt \lesssim$~30~GeV) the inclusive spectra of charged electrons or muons are 
dominated by decays of heavy-flavour hadrons. Using 1.13~pb$^{-1}$ of early 2010 data, for which 
non-isolated low-\pt~electrons were recorded without any HLT selection bias, electron and muon signals are 
extracted from the dominant background of QCD fakes, conversions (in the electron case) and 
pion and kaon decays in flight (in the muon case). The well-understood theoretical prediction for the
contribution from  $W$/$Z$/$\gamma^{\ast}$ production is then subtracted to leave the heavy-flavour component
of the signal. 

\begin{figure}
\resizebox{0.95\columnwidth}{!}{%
\includegraphics{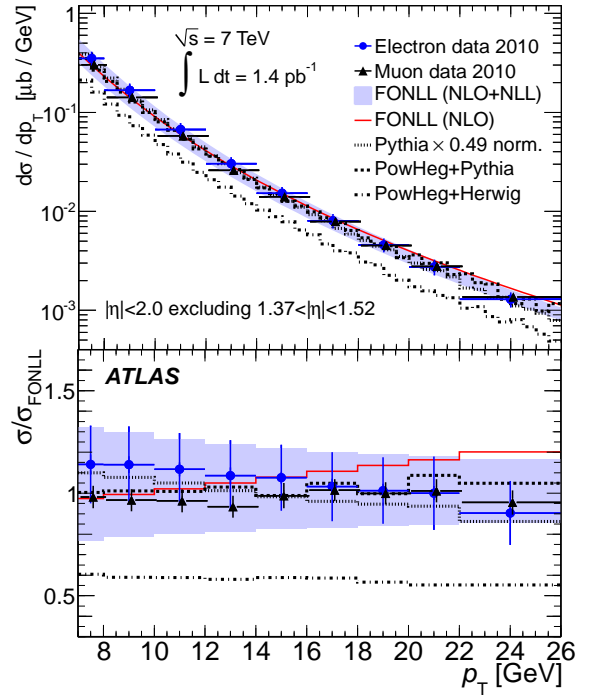} }
\caption{Electron and muon differential cross-sections from heavy-flavour production as a function of~\pt~for 
$|\eta| <2.0$  excluding $1.37 < |\eta| <1.52$.
The ratio of the measured and predicted cross-sections to the FONLL calculation is given in the bottom. The 
{\tt PYTHIA} (LO) cross-sections are normalised to the data.}
\label{fig:1}       
\end{figure}

The results~\cite{bib:incele} for electrons and muons within $7< \pt < 26$~GeV and
  $|\eta| <2.0$  (excluding $1.37 < |\eta| <1.52$) are shown in
Fig.~\ref{fig:1}, compared to the predictions from the FONLL theoretical framework, in which the heavy 
quark production cross-section is calculated in pQCD by matching the Fixed Order NLO terms with NLL 
high-\pt~resummation. Good agreement with the FONLL calculation is observed, including the sensitivity 
to the NLL terms.

\subsection{Quarkonium cross-section measurements}
\label{sec:2}

Despite being among the most studied of the bound quark systems, there is still no clear understanding
of the production mechanisms for quarkonium states like the $J/\psi$ and the $\Upsilon$ 
that can consistently explain both the
production cross-section and spin alignment measurements in $e^+ e^-$, hadron and heavy ion collisions. 
Data from the LHC allow tests of theoretical models of quarkonium production in a new 
energy regime.

The inclusive $J/\psi$ and  $\Upsilon(1S)$  production cross-sections are measured at ATLAS in the di-muon decay 
channel using 2.3~pb$^{-1}$ and 1.1~pb$^{-1}$ of 2010 data, respectively~\cite{bib:jsxs,bib:upxs}. The true numbers of $J/\psi$ or  
$\Upsilon(1S)$ candidates are recovered from the observed di-muon pairs by applying event weights to unfold the 
response of the detector, with the efficiency factors being determined almost entirely from the data. 
(Monte Carlo is used to provide trigger efficiencies in finer binning than would be
possible with the data statistics, with scale factors being obtained from data.) The $J/\psi$ and $\Upsilon(1S)$ yields are then determined from fits to the 
di-muon mass in bins of \pt~and $\eta$. 

\begin{figure}[h!]
\begin{tabular}{cc}
\resizebox{0.90\columnwidth}{!}{%
\includegraphics{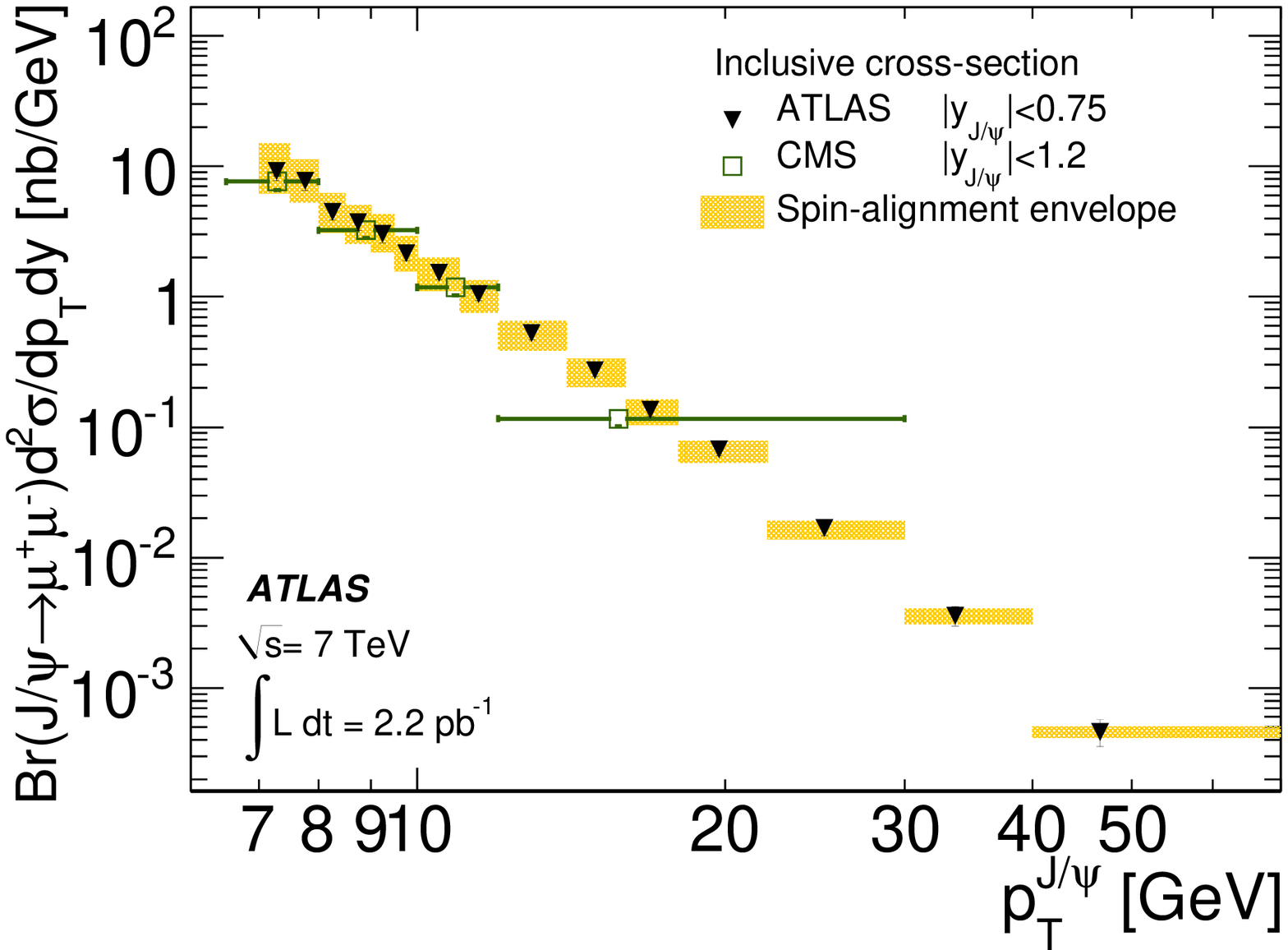} }\\
\resizebox{0.93\columnwidth}{!}{%
\includegraphics{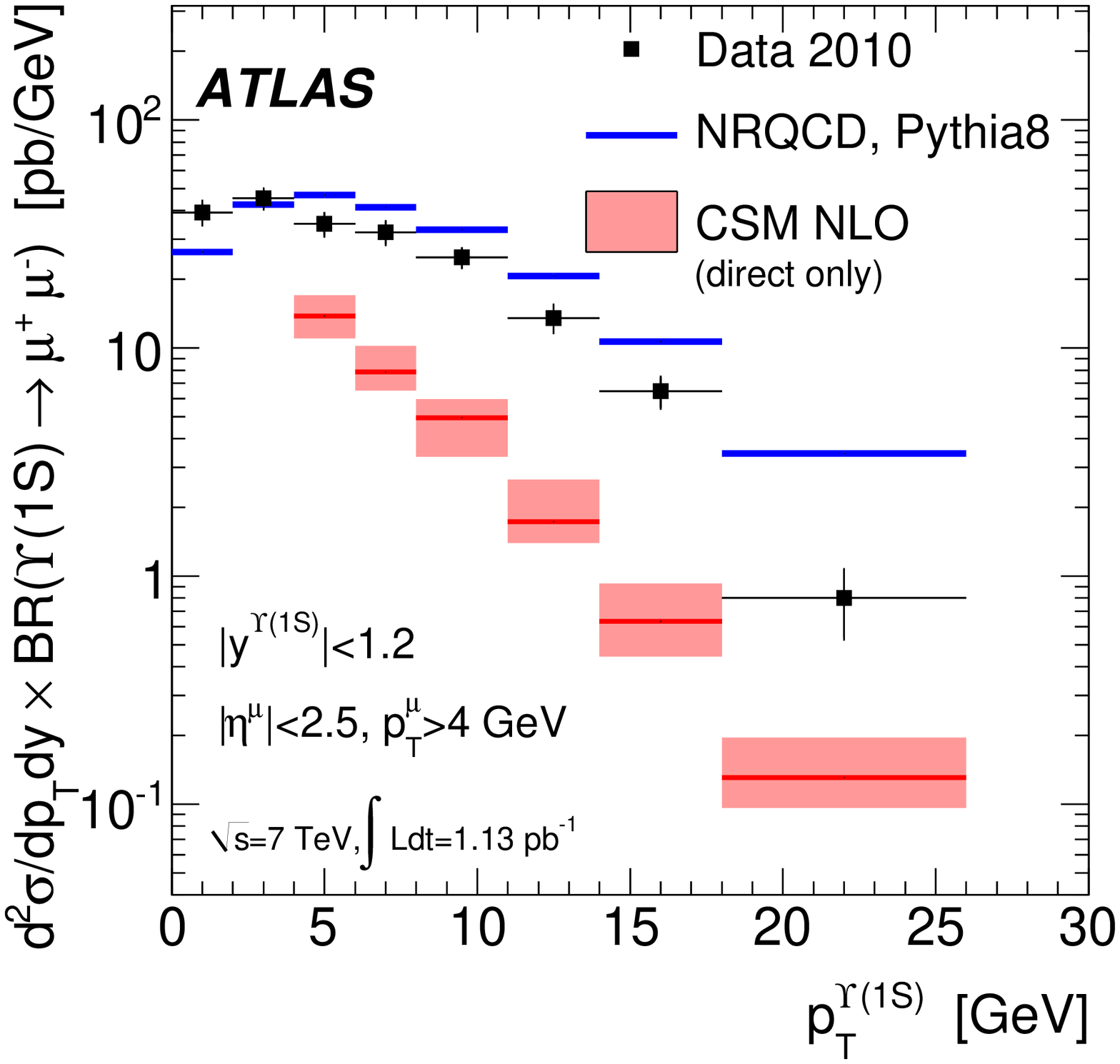} }
\end{tabular}
\caption{(Top) Inclusive $J/\psi$ production cross-section as a function of \pt~for the example  $J/\psi$ rapidity
bin of $|y|<0.75$, compared to the equivalent results from CMS.
(Bottom) Inclusive $\Upsilon(1S)$ production cross-section as a function of \pt~for the example   $\Upsilon(1S)$ 
rapidity bin $|y|<1.2$, for the given kinematic acceptance of the two muons. The result is compared to
the colour-singlet model (CSM) NLO prediction for which the shaded area shows the change in the
prediction when varying the renormalisation and factorisation scales (nominally $m_{\rm T}$) by a factor of two. 
The CSM NLO calculation accounts only for direct production of  $\Upsilon(1S)$ mesons and excludes 
any feed-down from excited states. 
The NRQCD prediction as implemented in PYTHIA8 is also shown for a particular choice of 
parameters. }
\label{fig:2}       
\end{figure}

In the $J/\psi$ analysis, a correction for the detector acceptance (the probability for muons to fall into 
the fiducial volume of detector) is also applied in order to obtain a total cross-section measurement. 
Since this correction depends on the unknown spin alignment of the $J/\psi$, an envelope of all possible 
polarisation assumptions is taken as an additional theoretical uncertainty. For the $\Upsilon(1S)$ measurement
no acceptance correction is applied, and the fiducial cross-section is reported in a restricted region of phase space 
where the muons are detected, free of any assumptions about the spin alignment. The  $J/\psi$ and  $\Upsilon(1S)$ 
 inclusive cross-sections in an example rapidity ($y$) bin are shown in Fig.~\ref{fig:2}. For the  $\Upsilon(1S)$
case, the results are compared to predictions from {\tt PYTHIA} 8.135 using the NRQCD framework and to
NLO QCD calculations as implemented in MCFM. Significant deviations from the data are seen in both
predictions, though MCFM does not include any feed-down from higher mass states 
which was estimated to contribute about a factor of two at the Tevatron.

\begin{figure}
\begin{tabular}{cc}
\resizebox{0.92\columnwidth}{!}{%
\includegraphics{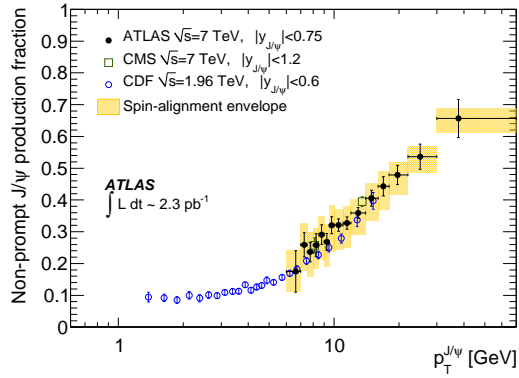} }\\
\resizebox{0.92\columnwidth}{!}{%
\includegraphics{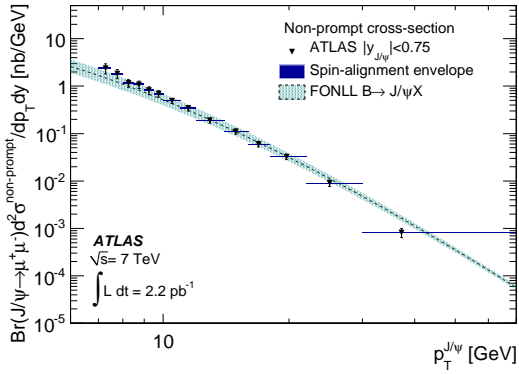} }\\
\resizebox{0.92\columnwidth}{!}{%
\includegraphics{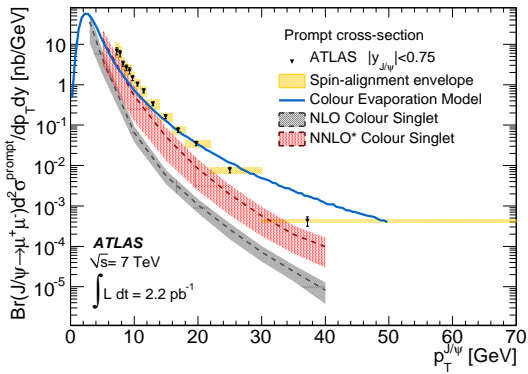} }
\end{tabular}
\caption{(Top) $J/\psi$ non-prompt to inclusive fraction as a function of the $J/\psi$ \pt~compared to equivalent 
results from CMS and CDF. (Middle) Non-prompt $J/\psi$ production cross-section as a function of $J/\psi$ 
\pt~, compared to predictions from FONLL. (Bottom) Prompt $J/\psi$ production cross-section as a function of
$J/\psi$ \pt, compared to predictions from NLO and NNLO$^\ast$  calculations, and the Colour Evaporation Model.
All plots are in the example $J/\psi$ rapidity bin  $|y|<0.75$.}
\label{fig:3}       
\end{figure}

Of the $J/\psi$ inclusive cross-section, the  non-prompt fraction coming from $B$-hadron decays can be
identified experimentally due to the associated displacement of the $J/\psi$ vertex in the transverse plane, 
$L_{xy}$, due to the long lifetime of the parent $B$-hadron.
A simultaneous unbinned maximum likelihood fit to the $J/\psi$ invariant mass and pseudo-proper decay time, 
$\tau = {L_{xy}m_{\rm{PDG}}^{J/\psi}}\,/\,{\pt^{J/\psi}}$, is used to extract this fraction 
from the data, allowing the prompt and non-prompt cross-sections to be obtained.
The results in an example rapidity bin are shown in Fig.~\ref{fig:3}. For the non-prompt component, good
agreement is seen with the predictions of FONLL. For the prompt component, the data are consistent with
NNLO$^{\ast}$ calculations.

\subsection{\boldmath{$D$} meson cross-section measurements}

\begin{figure}
\begin{tabular}{cc}
\resizebox{0.95\columnwidth}{!}{%
\includegraphics{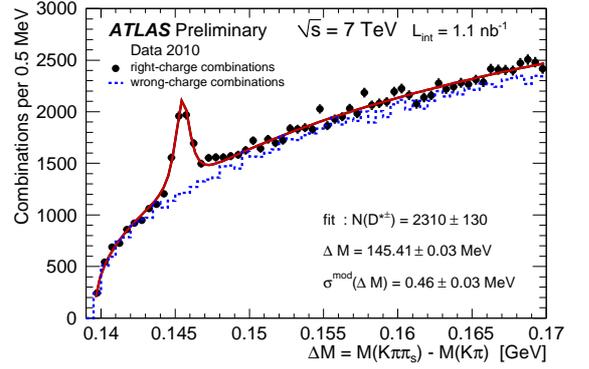} }\\
\resizebox{0.95\columnwidth}{!}{%
\includegraphics{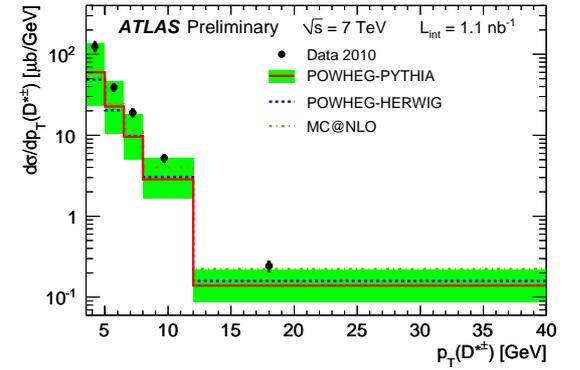} }
\end{tabular}
\caption{
(Top) The distribution of the mass difference, $\Delta M = M(K \pi \pi_s) - M (K \pi)$, for $D^{\ast \pm}$  
candidates.
The dashed lines show the distribution for wrong-charge combinations and the solid curves represent the 
fit results.
(Bottom) Differential cross-section for  $D^{\ast \pm}$ mesons as a function of \pt~for data  
compared to the NLO QCD calculations of POWHEG-PYTHIA, POWHEG-HERWIG and MC@NLO for $D$-mesons
produced within $|y| < 2.1$. The bands show the estimated theoretical uncertainty of the POWHEG-PYTHIA calculation.
}
\label{fig:4}       
\end{figure}

Using an integrated luminosity of 1.1~nb$^{-1}$,  $D^{\ast \pm}$, $D^\pm$ and $D_s^\pm$ charmed mesons
with $\pt > 3.5$~GeV and $|y| < 2.1$ are reconstructed using tracks measured in the ATLAS ID~\cite{bib:dxs}.
Taking the example of the $D^{\ast \pm}$, which is identified in the decay 
channel $D^{\ast \pm} \to D^0 \pi^{\pm}_s \to (K^- \pi^+)\,\pi^{\pm}_s$, where the $\pi^{\pm}_s$ is the slow
pion in the $D^{\ast \pm}$ decay frame, pairs of oppositely-charged tracks with $\pt > 1.0$~GeV
 are combined to form  $D^0$ candidates, with  kaon and pion masses assumed in turn 
for each track in order to calculate the invariant mass.
Any additional track, with $\pt > 0.25$~GeV and a charge opposite to that of the kaon track, is
assigned the pion mass and combined with the  $D^0$ candidate to form a $D^{\ast \pm}$ candidate.
A  clear signal is seen in Fig.~\ref{fig:4} (Top) in the distribution of the mass difference 
$\Delta M = M(K \pi \pi_s) - M (K \pi)$ at the nominal value of $M(D^{\ast \pm}) - M(D^0)$.
From a fit to the $\Delta M$ distribution, a  $D^{\ast \pm}$ yield of $2310 \pm 130$ is obtained, and 
its mass found to be $145.41 \pm 0.03$~MeV, in agreement with the PDG world average. Similarly for the  $D^\pm$  
(yield $1546 \pm 81$) and $D_S^\pm$  (yield $304 \pm 51$) mesons, masses consistent with PDG world averages are found. 

Using Monte Carlo to correct for the detector response, the $D^{\ast \pm}$ and $D^\pm$ production
cross-sections (in the kinematic acceptance for the $D$-mesons of $\pt > 3.5$~GeV and $|y| < 2.1$)
are found, as shown in in  Fig.~\ref{fig:4} (Bottom) for the  $D^{\ast \pm}$  example.
Within the large theoretical uncertainties, NLO QCD predictions are seen to be consistent with the data,
the uncertainty on which is dominated by the statistical error.

\section{Lifetime measurements}

\subsection{\boldmath{$B$}-hadron average lifetime}
\label{sec:3}

Precise measurements of $B$-hadron lifetimes allow tests of theoretical predictions from the Heavy Quark
Expansion framework, which can predict lifetime ratios for different $B$-hadron species with per cent
level accuracy. With the goal of preparing the way for precision measurements of $B$-hadron lifetimes, 
an average lifetime measurement of inclusive $B \to J/\psi X \to \mu\mu X$ is made on the full 2010 dataset
(35~pb$^{-1}$)~\cite{bib:avblife}.  With orders of magnitude higher statistics than for fully reconstructed exclusive $B$-mesons
 or $B$-baryon candidates, the inclusive sample enables a detailed investigation of the decay length resolution 
and the impact of the residual misalignment of the tracking system.

\begin{figure}
\begin{tabular}{cc}
\resizebox{1.0\columnwidth}{!}{%
\includegraphics{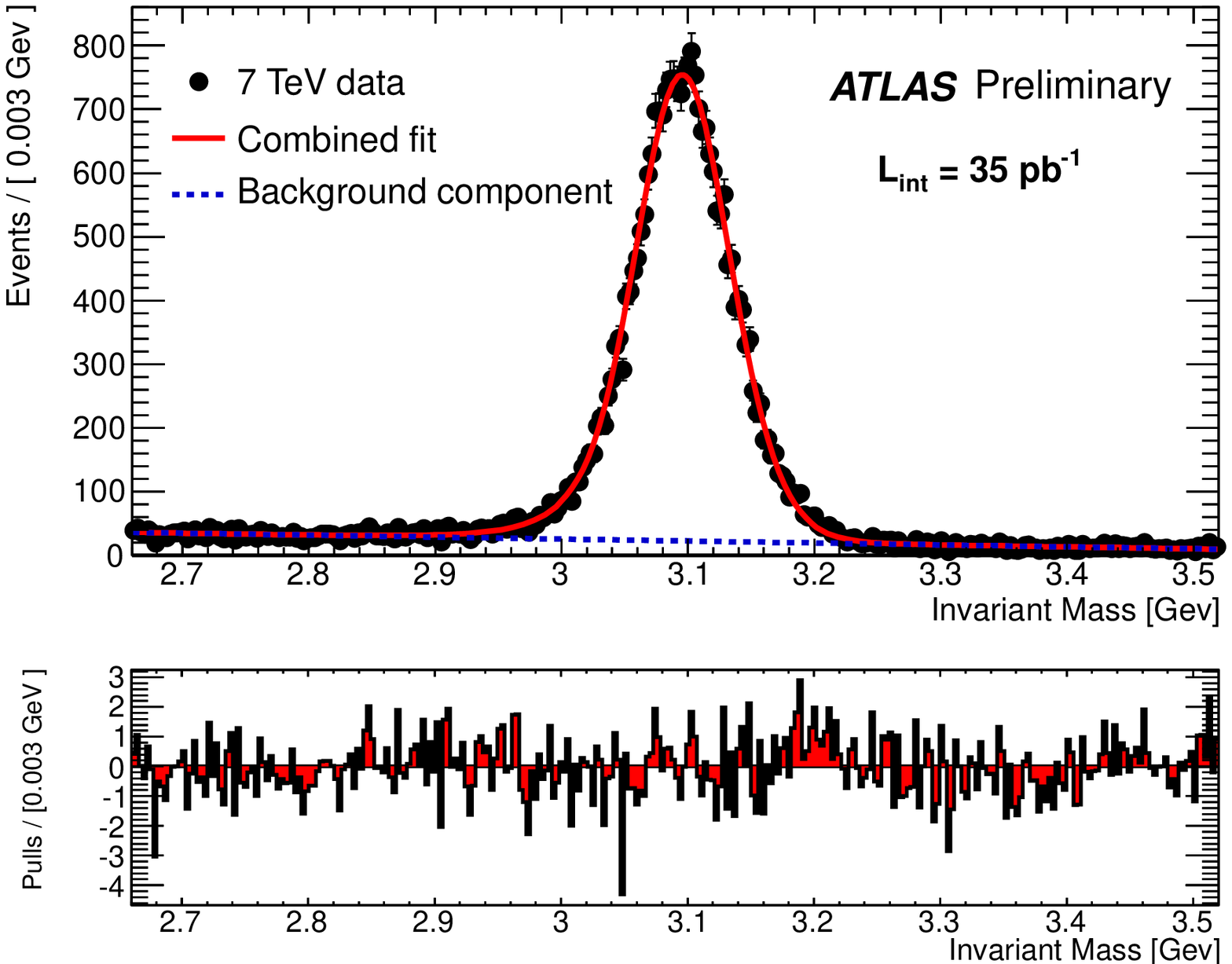} }\\
\resizebox{1.0\columnwidth}{!}{%
\includegraphics{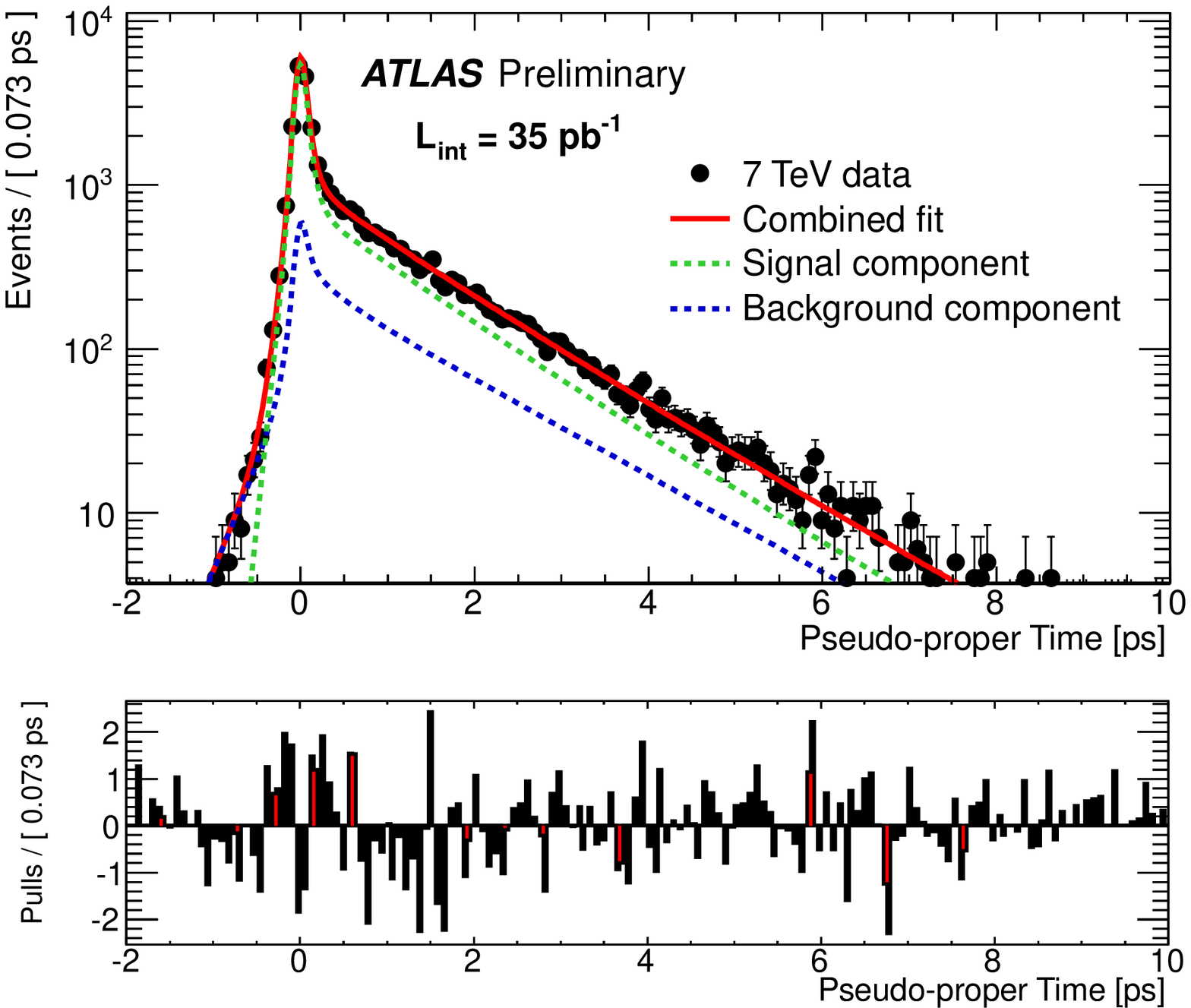} }
\end{tabular}
\caption{ Invariant mass (Top) and pseudo-proper decay time (Bottom) projections (with their respective pull distributions)
of the simultaneous fit to these distributions in the $B$-hadron average lifetime measurement, showing the data 
and the signal, background and combined fit results. }
\label{fig:5}       
\end{figure}

The inclusive lifetime measurement gives the average lifetime of the admixture of $B$-hadrons 
produced at the LHC and decaying to final states including a $J/\psi$. As discussed above, $J/\psi$
mesons produced from the decays of $B$-hadrons are non-prompt, having a displaced 
decay vertex due to the $B$-hadron lifetime. 
The average $B$-lifetime is thus extracted from the data by performing an unbinned maximum 
likelihood fit simultaneously to the $J/\psi$ invariant mass and the pseudo-proper decay time.
In order to extract the real lifetime of the $B$-hadrons, which are not fully reconstructed,
a correction for the smearing introduced by the use of  the pseudo-proper decay time in the fit is required.
This correction (``$F$-factor'') is obtained from Monte Carlo, with the  $J/\psi$ spectrum in $B$-hadron
decays re-weighted to match BaBar data.
The invariant mass and pseudo-proper decay time projections of the fit, with the respective
pull distributions, are shown in Fig.~\ref{fig:5}. 
The average $B$-lifetime is measured to be:
\begin{equation}
\langle {\tau}_B \rangle = 1.1489 \pm 0.016~(\rm{stat.}) \pm 0.043~(\rm{syst.})~ps,
\end{equation}
with the main systematic uncertainty originating from the uncertainty (in 2010) on the radial 
alignment of the ID.
This result is in agreement with the most recent one from the CDF collaboration and the 
expected average lifetime computed using PDG lifetimes and production fractions from different $B$-hadron species.

\subsection{{\boldmath{$B^0_d$}} and {\boldmath{$B^0_s$}} lifetimes}

The lifetimes of $B^0_d$ and $B^0_s$ mesons are determined from their exclusive decay modes
 $B_d^0 \to J/\psi K^{\ast 0}$ and  $B_s^0 \to J/\psi \phi$, using the $J/\psi$ decay to a di-muon final state,
based on an integrated luminosity of 40~pb$^{-1}$~\cite{bib:blife}.
The study of the $B_s^0 \to J/\psi \phi$ decay is of special interest as it allows the measurement of the $B^0_s$ 
mixing phase which can generate CP violation in this channel. The SM prediction for this CP violating
phase is small meaning that any measured excess would be a clear indication of new physics.
In addition the light ($B_L$) and heavy ($B_H$) mass eigenstates have two distinct decay
widths $\Gamma_L$ and $\Gamma_H$ which have been determined at the Tevatron using a 
technique of time-dependent angular analyses that allow the simultaneous extraction of the 
CP-even and CP-odd amplitudes. 
The  $B^0_d$ channel provides a valuable testing ground for measurements
of the $B^0_s$ decay  due to the equivalent event topology with advantage of higher statistics.

The  $K^{\ast 0}$ or  $\phi$ candidates are reconstructed by selecting all pairs of oppositely charged tracks
not previously identified as muons, fitted to a common vertex with the two muon tracks from the  $J/\psi$.
Quadruplets of tracks passing certain selections then fully reconstruct the $B^0_d$ or $B^0_s$
candidates. Since the $B$-hadron is fully reconstructed, the  proper decay time
 $\tau = {{L_{xy}{m_{B}}}}\,/\,{\pt{_B}}$, can be used in an unbinned maximum likelihood fit 
together with the  invariant mass, in order to extract the yield, mass and lifetime in each channel.

For $B^0_d$ the lifetime is found to be
\begin{equation}
{\tau}_{B_d^0} = 1.51 \pm 0.04~(\rm{stat.}) \pm 0.04~(\rm{syst.})~ps
\end{equation}
and for  $B^0_s$
\begin{equation}
{\tau}_{B_s^0} = 1.41 \pm 0.08~(\rm{stat.}) \pm 0.05~(\rm{syst.})~ps,
\end{equation}
where a main source of systematic uncertainty comes from the detector alignment. 
With a limited number of $B^0_s$ candidates ($463 \pm 26$) the current measurement ignores the lifetime
difference of the mass eigenstates. However, within the current precision the fitted value agrees well with 
the world average of the average lifetime ($1.472 \pm 0.026$~ps).

\section{Summary}

The ATLAS heavy-flavour physics programme has made many important measurements of production cross-sections
and masses and lifetimes of various heavy-flavour states. In several cases these are already challenging 
current theoretical models at the LHC energy regime. Moreover, the demonstration of the experimental techniques 
paves the way for future measurements and searches for CP violation and rare decays. The ability to perform
these measurements has been particularly dependent on the excellent performance of the
 ATLAS ID, MS and trigger systems.

%
%

\end{document}